# One-dimensional Multiferroic Semiconductor WOI$_3$: Unconventional Anisotropic $d^1$ Rule and Bulk Photovoltaic Effect

Zhihao Gong[1#], Yechen Xun[2#], Zhuang Qian[3], Kai Chang[1], Jingshan Qi[4*], Hua Wang[1*]

[1]ZJU-Hangzhou Global Scientific and Technological Innovation Center, School of Physics, Zhejiang University, Hangzhou 311215, China

[2]Department of Physics and Astronomy,

The University of Tennessee, Knoxville, Tennessee 37996, USA

[3]Institute of Natural Sciences,

Westlake Institute for Advanced Study, Hangzhou, Zhejiang 810024, China

[4]Tianjin Key Laboratory of Quantum Optics and Intelligent Photonics, School of Science,

Tianjin University of Technology, Tianjin 300384, China

*Correspondence to: qijingshan@email.tjut.edu.cn, daodaohw@zju.edu.cn

## Abstract

The pursuit of multiferroic magnetoelectrics, combining simultaneous ferroelectric and magnetic orders, remains a central focus in condensed matter physics. Here we report the centrosymmetric, one-dimensional (1D) antiferromagnetic WOI$_3$ undergoes a strain-induced ferroelectric distortion. The paraelectric-ferroelectric transition is originated from the unconventional anisotropic $d^1$ mechanism, where an unpaired $d$ electron of each W$^{5+}$ ion contributes to magnetic orders. Employing a Heisenberg model with Dzyaloshinskii–Moriya interaction, we predict an antiferromagnetic spin configuration as the paraelectric ground state, transitioning to a ferroelectric phase with noncollinear spin arrangement under uniaxial strain. The ferroelectric polarization and noncollinear spin arrangement can be manipulated by varying the applied strain. While the energy barriers for switching ferroelectric polarizations with magnetic orders are on the order of a few dozen of meV, the shift current bulk photovoltaic effect (BPVE) exhibits remarkable differences, providing a precise and valuable tool for experimentally probing the interplay of ferroelectric and magnetic orders in 1D WOI$_3$.

# I. Introduction

Multiferroics have been a focal point in the realm of condensed matter physics [1–6], and are promising to be vastly applied in nonvolatile data storage, sensors, actuators and so on [6], especially for those materials combining ferroelectricity (FE) and magnetic behaviors [4]. Due to clean surfaces and large dielectric constant in miniaturization of the electronic devices [6], low-dimensional (1D-, 2D-) multiferroic materials have attracted significant attentions for these days. Based on the mechanism of generating multiferroicity [2,5], multiferroic materials can be classified into Type-I (sources of FE and magnetism are independent) and Type-II (magnetism causes FE) categories. Notable examples of Type-I materials include artificial low-dimensional materials, like halogen-decorated phosphorene bilayers [7], 2D Van der Waals (vdW) heterostructure of $Cr_2Ge_2Te_6$ onto $In_2Se_3$ [8], etc. and transition-metal halide monolayers, like $(CrBr_3)_2Li$ [9], while $Hf_2VC_2F_2$ MXene Monolayer [10] and few-layer $NiI_2$ [11] belong to Type-II.

Type-I multiferroic materials violating the conventional $d^0$ rule of octahedral distortion for FE by external strain engineering [12,13], and intrinsic anisotropic $d^1$ rule [14–16] provide the potential for simultaneous magnetic and ferroelectric ordering. Increasing the lattice constants, epitaxial strains and chemical pressure are discussed in stabilizing the FE state in antiferromagnetic (AFM) materials like $SrMnO_3$ [12] and $BaMnO_3$ [13], respectively. On the other side, the materials of $d^1$ configurations, if governed by Hund's rule, exhibit no pseudo Jahn-Teller effect (PJTE) on dipolar distortion and no ferroelectric instability [17,18], due to the mismatch in spin multiplicities of the lowest unoccupied molecular orbital (LUMO) and the highest occupied molecular orbital (HOMO). Nevertheless, $d^1$ configurations haven't been fully excluded from satisfying the condition of PJTE [18]. The anisotropic $d^1$ rule is recently proposed in predicting the ground-state FE and AFM orders in the 2D $VOX_2$ (X=Br, Cl, I) monolayer [14]. Distinct from the conventional $d^1$ configurations, the anisotropy of the local octahedral field could lead PJTE instability by splitting the threefold degenerate $t_{2g}^*$ molecular orbitals. Additionally, the spiral-spin noncollinear magnetic order is then predicted by introducing the effect of the Dzyaloshinskii–Moriya interaction (DMI) [16].

In this work, we report the strain-induced coexistence of ferroelectricity and antiferromagnetism of the 1D vdW semiconductor $WOI_3$ based on density functional theory (DFT) calculations. Distinct from the octahedral quasi-1D $WOX_4$ (X=Br, Cl, I) [19] and 1D $WOF_4$ [20], in which the intrinsic ferroelectricity is rooted from the electronic $d^0$ configurations, the 1D $WOI_3$ material exhibits anisotropic $d^1$ configuration despite sharing similar octahedral structures, and is predicted to experience a paraelectric-ferroelectric phase transition under axial tensile strain. Furthermore, the ground-state AFM order is predicted, and the emerging DMI effect, caused by strain-induced FE distortion, could alter the magnetic orientations from collinear configurations to non-collinear configurations. Both the FE polarizations and non-collinearity are monotonically increased with increasing strain. We further explore the shift current bulk photovoltaic effects (BPVE), which are recognized as the second-order nonlinear optical responses, under varying strains and different ferroelectric and magnetic states in the 1D $WOI_3$. The shift current is one of the direct BPVE photocurrents regained massive attention in both theoretical [21–27] and experimental [28–31] condensed matter physics due to its geometric nature and potential applications [32–34]. It is noteworthy that the synthesis of bulk vdW WOI$_3$ [35] and other MYX$_3$ (M=Mo, W; Y=O, S, Se; X=Cl, Br) [36–38], sharing the same crystal structure, have been reported decades ago. Encouraged by the experimental achievements, this work suggests the 1D semiconductor WOI$_3$ as a promising multi-tunable platform for nonlinear optics, paving the way for next-generation devices that harness the intricate interplay of electricity and magnetism through nonlinear optical responses.

## II. Methods

Density functional theory (DFT) [39,40] calculations are performed by Vienna *Ab initio* Simulation Package (VASP) [41,42], which employs projected augmented-wave (PAW) approach for the core electrons. The generalized gradient approximated (GGA) functionals [43,44] of Perdew-Burke-Ernzerhof (PBE) type parameterization [45] are implemented to consider the exchange-correlation energy of all valence electrons. Due to strong electronic localization for $5d$ electrons of W atoms, the Hubbard correction with simplification of Dudarev *et. al* [46] is further added to the GGA functionals with parameters, $U_{\text{eff}} = U - J = 2$ eV. The $1 \times 1 \times 2$ supercell of 1D WOI$_3$ material is mainly used in DFT calculations, allowing for six different magnetic configurations being taken into account. A cutoff energy for the plane-wave basis sets is set as 450 eV, and a $1 \times 1 \times 7$ $k$ point sampling is used. The DFT results are consistently obtained under the convergence criteria of $0.01$ eV/Å and $1 \times 10^{-6}$ eV for residual forces in ion relaxations and energy difference between successive self-consistent field (SCF) calculation steps, respectively.

Based on DFT calculations, following methods are further conducted to investigate the FE and magnetic orders of WOI$_3$. The standard Berry phase approach [47] is employed to calculate FE polarizations. The torque method [48,49] is applied for magnetic anisotropy energy (MAE) profile with an error tolerance of $\sim 0.01\pi$ solid angle. In the Monte Carlo (MC) simulations of the Heisenberg model for WOI$_3$, the orientations of magnetic moment are sampled in a supercell of $1 \times 1 \times 2000$ grids to get rid of interference from the boundaries. The MC simulations are performed under a temperature of $10^{-8}$ K, and the Metropolis algorithm [50] and the Hinzke-Nowak methods [51] are introduced to improving the sampling efficiency. Initial magnetic configurations are deployed randomly, and a dynamic balance of the magnetic moments is reached after $1 \times 10^5$ steps.

The Wannier-interpolation scheme [52,53] is employed in calculations for shift current responses. The maximally-localized Wannier functions (MLWF) are obtained iteratively with initial guess constructed by projection functions of W $d$-, I $p$-, and O $p$-orbitals, as implemented in the Wannier90 code [54]. The tight-binding model is built from the basis set of MLWF, and the Wannier Berri code [55] is then performed to calculate the shift current response. Within the calculations, a dense $k$-point mesh of $1 \times 1 \times 400$ for integrations and a fixed width of $0.025$ eV for broadening Dirac $\delta$-functions are adapted.

The visualization of the crystal structures are carried out by VESTA software [56]. The output data from above codes are in units of 3D systems. The 1D results for shift current response (nm$^2 \cdot$ μA/V$^2$), linear optical response (pS $\cdot$ cm), and FE polarization (pC $\cdot$ m) are then obtained by multiplying data with the cross section of $25$ Å $\times 25$ Å.

# III. Results

## A. Strain-induced ferroelectricity

We start with a brief description about the WOI$_3$ crystal in the paraelectric (PE) phase. The ladder-like 1D WOI$_3$ is formed by two 1D chains. Within the chains, highlighted in Fig. 1(a), each tungsten (W) atom is confined in an irregular octahedron four planer iodine atoms (I) and two apical oxygen atoms (O). A $1 \times 1 \times 2$ cell of WOI$_3$ is used in DFT calculations (Fig. 1(b)). The crystal structure is built with inserting $17.2$ Å and $21.0$ Å vacuum layers over $a$ and $b$ directions, and is then optimized with a relaxed lattice constant $c = 3.825$ Å in GGA-level accuracies. The PE structure of WOI$_3$ is in the space groups $Pmmm$ (No. 47) with two W atoms residing in the plane composed of the six planar I atoms. The calculated energy of extracting 1D WOI$_3$ wires from the bulk WOI$_3$ is approximately $25$ meV/atom [57], implying experimental feasibility for exfoliation, akin to the exfoliation energy ($\sim 52$ meV/atom) of a 2D graphene sheet.

Applying uniaxial tensile strain $\varepsilon$, the WOI$_3$ crystal demonstrates the capability to undergo a PE-to-FE structural phase transition. The nudged elastic band (NEB) calculations [58,59] under $\varepsilon = 3\%$ yield two adiabatic energy pathways, and the results show that the structure in the FE/FE$_{anti}$ phase is in the ground state (Fig. 1(c)), while the AFE/AFE$_{anti}$ structure is metastable (Fig. 1(d)). The former belongs to $Pmm2$ (No. 25) space group with W$^{5+}$ ions for both chains shifting along the same $c$ direction, and the latter is in $Pm/2$ (No. 31) space group with cancelled polarization due to opposite $c$-direction displacements of W$^{5+}$. The energy barrier of reversing the FE polarizations in WOI$_3$ typically ranges in order of several dozens of meV/u.c.. Specifically, the barrier of PE state is $31.2$ meV/u.c. and barrier between FE and AFE is $16.2$ meV/u.c. under strain of $\varepsilon = 3\%$ (Figs. 1(c)-(d)). Compared with other low-dimensional materials, like 2D $\alpha$-In$_2$Se$_3$ ($\sim 850$ meV/u.c.) [60], the FE switching in WOI$_3$ can be easily achieved by applying electric fields.

For varying strain intensity $\varepsilon$, the transitions between PE, AFE and FE phases are elucidated by examining the energy differences, $\Delta E = E - E_{\text{PE},0\%}$ (depicted in Fig. 1(e) and summarized in Table. I) along with the specific angle $\alpha$ (Table. I). Here, the angle $\alpha$ refers to the angle at lower left corner of the cage shown in Fig. 1(b). In the PE state under increasing intensity of strain $\varepsilon$, both $\Delta E_{\text{PE}}$ and $\alpha_{\text{PE}}$ become larger. Under strains of $2\% \leq \varepsilon \leq 5\%$, the differences between $\Delta E_{\text{FE}}$ ($\Delta E_{\text{AFE}}$) and $\Delta E_{\text{PE}}$ are significant, and the FE state becomes the lowest-energy state with $\Delta E_{\text{FE}}$ being smaller than $\Delta E_{\text{AFE}}$ by a few meV. Regarding structural distortion, $\alpha_{\text{FE}}$ for FE state are consistently smaller than $\alpha_{\text{PE}}$, while $\alpha_{\text{AFE}}$ for AFE state are larger than $\alpha_{\text{PE}}$. With increasing strength of strains, the differences in angles become larger. However, for strains $\varepsilon < 1.8\%$, both $\alpha$ and $\Delta E$ of FE, AFE, and PE structures exhibit virtually no difference, which indicates that the PE-to-FE critical point (CP) may locate near strain $\varepsilon = 1.8\%$. The phase transition is further evident in the FE polarizations depicted in Fig. 1(f), which increases monotonically above the CP strain and vanishes below it. Our prediction of the strain-induced FE phase transition in WOI$_3$ appears to be highly feasible for experimental validation by hysteresis loop measurements, particularly in light of the spontaneous structure distortions derived in recent X-ray diffractogram of 1D MoOBr$_3$ [38].

## B. Anisotropic $d^1$ mechanism

The emergence of ferroelectricity in the strained 1D WOI$_3$ material could be attributed to the anisotropic $d^1$ rule. This rule has been recently introduced to elucidate potential multiferroic properties in 2D VOX$_2$ (X=Cl, Br, I) monolayer. Analysis of spatial charge densities and projected densities of states (PDOS) are conducted, as detailed in References [14–16]. Due to an unpair $d$ electron of W$^{5+}$ centered at each octahedron, the WOI$_3$ is a non-$d^0$ system. We substantiate the applicability of the anisotropic $d^1$ rule for the strain-induced FE transition in WOI$_3$ by examining the charge density (Fig. 2(a)) and the PDOS (Figs. 2(b) and S2-S4 in Ref. [57]). The $xy$-coordinates are clearly marked in Fig. 2(a), and additional information regarding the selected $xy$-coordinates and spin polarizations can be found in Ref. [57]. The degree of coupling between W $d$-orbital and the O $p$-orbital is elucidated from the spin-polarized PDOS of ($\varepsilon = 0\%$, PE), ($\varepsilon = 3\%$, PE & FE), and ($\varepsilon = 5\%$, PE & FE). With a discernible level of hybridization for $d_{xz}(d_{yz})$-orbitals and O $p$-orbital, the coupling between $d_{xy}$-orbital and the O $p$-orbital remains negligible. The unpaired electrons occupying the $d_{xy}$-orbital (Fig. 2(a)) do not cause significant hindrance when W$^{5+}$ moves along the W-O chain. On the other hand, the spin multiplicity of HOMO and LUMO configurations is illustrated from the PDOS of spin-up unit cell (Fig. 2(b)). Near the Fermi energy, the spin-up PDOS of occupied $d_{xy}$ orbital indicates the HOMO configuration as $(t_{1u})^6(d_{xy}\uparrow)^1$. The observable spin-down PDOS of unoccupied $d_{xz}$ ($d_{yz}$) orbitals informs a possible configuration of $(t_{1u})^5(d_{xy}\uparrow)^1(d_{yz}\downarrow + d_{xz}\downarrow)^1$ for LUMO, which share the same spin multiplicity as HOMO.

With hindrance-free $d_{xy}$ electrons and same spin multiplicities, the anisotropic $d^1$ configuration bears resemblance to the conventional $d^0$ configuration, and the HOMO and LUMO can hybridize under the polar displacement of W$^{5+}$ along the W-O bonds leading finite vibronic coupling, $F \neq 0$. The occurrence of PJTE instability depends on whether the vibronic-coupling contribution, $K_V \approx -2|F|^2/(\Delta E_{\text{LUMO-HOMO}})$, is sufficient enough to overcome the lattice stiffness, $K_0$. In the pristine WOI$_3$ ($\varepsilon = 0\%$), vibronic-coupling contribution might not be dominate due a large HOMO-LUMO gap, $\Delta E_{\text{LUMO-HOMO}}$, and the ground-state structure is in PE under $K_0 + K_V > 0$. When applying tensile strains ($\varepsilon > 1.8\%$) to elongate the W-O bonds, the decrease of $\Delta E_{\text{LUMO-HOMO}}$, implied by the reduction of the energy gap of ($\varepsilon = 3\%$, PE) (Fig. 2(b)), leads an augmentation for vibronic-coupling contribution, $K_V$. The FE polarization can then occur triggered by the PJTE instability with $K_0 + K_V < 0$. The HOMO state of distorted WOI$_3$, derived from the hybridization of HOMO and LUMO states of ($\varepsilon = 3\%$, PE), is eventually restabilized with an enlarged HOMO-LUMO gap, which is confirmed by the expanded energy gap of ($\varepsilon = 3\%$, FE) (Fig. 2(b)).

## C. Strain-tuning noncollinear antiferromagnetism

Next, we turn to the magnetic property of WOI$_3$ arising from the extra unpaired $d$ electrons of W$^{5+}$. To explore the influence from the strain-induced FE distortion onto ground-state magnetic configurations, we include spin-orbit couplings (SOC) in DFT calculations, and then conduct MC simulations ("Methods") based on the effective Heisenberg model [61].

In the Heisenberg model, effective magnetic moments of atoms are incorporated for the study of macroscopic magnetic order. Both the magnitudes and directional dependence of atomic magnetic moments are carefully checked as follows to ensure a rational modelling. The magnitude of magnetic moment of W$^{5+}$ is approximately $1.10\ \mu_B$, obtained by integrating spin charge density within the effective radius in VASP. The magnetic moments of I$^-$ and O$^{2-}$ range within $0.01$-$0.04\ \mu_B$, varying with distance from W$^{5+}$. These values align with the expected magnetic moments, considering the presence of one extra unpaired $d$ electron of W$^{5+}$ and four spin pairings of I$^-$ and O$^{2-}$, with small deviations due to covalent bonding. The magnetic moments remain consistent under different strain magnitude in our DFT calculations. According to the Mermin-Wagner theorem [62], directional isotropy for spins causes strong fluctuations to prohibit spontaneous symmetry broken in 1D and 2D spin systems. In magnetic ordered materials, like CrSbSe$_3$ studied in our previous work [63], magnetic anisotropy plays a key role in suppression of such spin fluctuations. To demonstrate magnetic anisotropy of WOI$_3$, we perform MAE calculations for 3D energy profile by rotating the atomic magnetic moments over entire space under $\varepsilon = 0\%$ (Fig. 3(a)), and maintain opposing magnetic moments in neighboring unit cells (labelled as AFM 1, see Fig. 3(c)). The substantial magnetic anisotropy of WOI$_3$ is evident from the MAE profile. Thus, within the Heisenberg model, the strength of magnetic moments localized at W sites is treated as constant, and magnetic anisotropy is considered. Additionally, we determine the energetically favorable easy axis by careful numerical examinations with an error tolerance of $0.01\pi$ in solid angle, which lies at an angle of $43.20°$ with respect to *a*-axis and $62.61°$ for *c*-axis. The hard axis stays at angles of $(230.40°, 32.86°)$. These spin orientations hold for various strains less than $\varepsilon = 10\%$, as confirmed by DFT calculations with the same error tolerance, demonstrating the stability of the easy axis orientation across different strains.

Based on the aforementioned insights, we introduce the general Heisenberg Hamiltonian

$$H = \sum_{i \neq j} \sum_{\alpha=(a,b,c)} \left[ J_{i,j;\alpha} s_i^\alpha s_j^\alpha + d_{i,j;\alpha}(\mathbf{s}_i \times \mathbf{s}_j)^\alpha \right] + \sum_i \sum_{\alpha=(a,b)} D_{i;\alpha}(s_i^\alpha)^2, \qquad (1)$$

governing the collective behavior of the fixed-strength magnetic moments at W$^{5+}$ sites, denoted by $\mathbf{s}_i = \{s_i^\alpha\}(\alpha = a, b, c)$. The magnetic anisotropy is addressed by the subscript $\alpha$ for all the coefficients in Eq. (1), including single-ion anisotropy coefficient, $D_{i;\alpha=(a,b)}$. For the PE structure (space group *Pmmm*), the asymmetric DMI of magnetic moments of 1D WOI$_3$ is zero due to inversion symmetry. However, when strain $\varepsilon(> 1.8\%)$ is applied, the FE deformation leads inversion symmetry breaking. Within a unit cell, a mirror plane includes both W$^{5+}$'s and transect the W-I$_2$-W structure along the *ac* plane. According to Moriya's rules [64,65], finite DMI could be expected for FE structure, with the DMI coefficient vector being along the *b*-direction, $d_{i,j;\alpha=b}$. To estimate the strain effect onto magnetic noncollinearity, the DMI term with $d_{i,j;\alpha}$, is included in Eq. (1) for each pairing sites of $\{i, j\}$. Here, $J_{i,j;\alpha}$ represents the anisotropic exchange coefficient.

Reasonable simplifications and truncations are further introduced in Eq. (1) to make MC simulations numerically feasible. The subscript $i$ of the single-ion anisotropy coefficients $D_{i;\alpha}$ is omitted by assuming homogeneity for all $W^{5+}$ sites, and the coefficients are collectively denoted as $\boldsymbol{D}$. As shown in Fig. 3(b), the nearest-neighboring (NN) terms of Eq. (1) are mainly considered with NN exchange coefficient vectors, $\{\boldsymbol{J_1}, \boldsymbol{J_2}\}$, and NN DMI coefficient vectors, $\{\boldsymbol{d_1}, \boldsymbol{d_2}\}$. An additional next-nearest-neighboring (NNN) exchange term with $\boldsymbol{J_3}$ is also included to account for the minority of longer-range terms.

The coefficients $\boldsymbol{J}_{k(=1,2,3)}$, $\boldsymbol{d}_{k(=1,2)}$, and $\boldsymbol{D}$ are determined by six kinds of specific magnetic configurations (Fig.3(c)). Among these configurations, there are four collinear configurations, including FM, AFM 1, AFM 2, and AFM 3. Two additional noncollinear configurations are with clockwise (CW) and anti-clockwise (ACW) orientations. The total energy from DFT calculations for the six configurations are listed with different magnetic anisotropy ($\alpha = a, b, c$) and different strains $\varepsilon$ in Table. S1 of Ref. [57] For each anisotropic direction $\alpha$, the determination of $J_{k(=1,2,3);\alpha}$ and $D_\alpha$ are involved with the linear equations parameterized the vectors of $s_i^\alpha s_j^\alpha$ and $(s_i^\alpha)^2$ (first column in Table. S1 of Ref. [57]) and their corresponding DFT energies plugged into Eq. (1). The values of $\boldsymbol{d}_{k(=1,2)}$ are subsequently determined by solving linear equations built from the remaining two noncollinear configurations. The determination process is repeated for various strains $\varepsilon (\leq 10\%)$. The NN parameters are listed in Table. II, and the results for $\boldsymbol{J_3}$ and $\boldsymbol{D}$ can be found in Table. S2 [57]. Given that the values of NNN $\boldsymbol{J_3}$ are approximately 2 orders of magnitude smaller than NN $\boldsymbol{J}_{k=1,2}$, the effects of exchange interactions from longer-than-NN sites are expected to be minor. Notably, $d_{2;b}$ is finite, while other NN DMI coefficients remain negligible under different strains $\varepsilon (\geq 3\%)$, which agrees with the predictions from Moriya's rules.

For the collinear magnetic configuration under each strain $\varepsilon$, the AFM 1 configuration is suggested to be the lowest-energy magnetic order, due to the positive $\boldsymbol{J_1}$ and negative $\boldsymbol{J_2}$. As the FE distortion intensifies, the enhancement of noncollinearity is implied from the increase of the DMI $d_{2;b}$. To manifest the impact of noncollinearity within the 1D crystal model, the MC simulations based on Eq. (1) are then conducted to investigate the ground-state noncollinear magnetic configurations. According to the simulation results, the parallel magnetic moment vectors in $l$-th unit cell within PE structure are expanded with an intersection angle, $\theta_l$, along the $a$-direction under the effect of FE distortion (as schematically shown in Fig. 3(d)). The growing noncollinearity with increasing strain $\varepsilon$ is quantitatively characterized by the statistical average angle, $\theta = \langle \theta_l \rangle$, which is listed in Table. II.

Summarizing the complex phase behaviors around the CP strain $\varepsilon = 1.8\%$, we present a semi-schematic phase diagram delineating FE, AFE and PE states in Ref. [57]. Each phase is highlighted by discernible differences in the structural distortion, the FE polarization, and the magnetic noncollinearity. It is essential to note that this phase diagram representation is just one of the several possible configurations. Establishing of a more accurate phase diagram would require sophisticated calculations and even experimental measurements, which lie beyond the scope of this work.

## D. Strain-induced shift current BPVE

Finally, we demonstrate the nonlinear optical (NLO) responses, specifically the shift currents BPVE under varying strain $\varepsilon$ and different ferroelectric and magnetic states for the 1D $WOI_3$ material. As depicted in the schematic diagram (Fig. 4(a)), these three key factors in tuning the shift current response can be utilized in the future 1D $WOI_3$-based NLO devices.

The formulation of the NLO response for shift current is briefly summarized here, with detailed derivations available in Refs. [21–24]. In this work, we focus on the $z$-direction shift current induced by $z$-polarized light, expressed as $J_{\text{sh};z} = 2\sigma^{z;zz}(\omega)E(\omega)E(-\omega)$, and the shift current conductivity $\sigma^{z;zz}(\omega)$ is formulated as follows

$$\sigma^{z;zz}(\omega) = -\frac{i\pi e^3}{2\hbar^2} \int_{BZ} \left(\frac{dk}{2\pi}\right) \sum_{n \neq m} f_{mn} \text{Im}(r_{nm;z} r_{mn}) \delta(\omega - \omega_{nm}). \tag{2}$$

Here, the difference for the Dirac-Fermi distributions and energies between of bands $n$ and $m$ are denoted by $f_{nm}$ and $\omega_{nm}$, respectively, and the summation in Eq. (2) runs over pairs of valence and conduction bands. For each $(n, m)$ bands, the dipole moment is given by the inter-band Berry connections, $r_{nm}(k) = i\langle n|\partial_k|m\rangle$, and its covariant derivative is defined as $r_{nm;z}(k) = \partial_k r_{nm} - i[A_n(k) - A_m(k)]r_{nm}$. The intra-band Berry connection of $n$ band is denoted as $A_n = i\langle n|\partial_k|n\rangle$. Here, $|n\rangle$ and $|m\rangle$ denote the cell-periodic Bloch states.

Based on the calculations of Eq. (2), the impact from strain engineering and FE distortion on the shift current response is explored (Figs. 4(b)). While maintaining $WOI_3$ in the FE and collinear AFM 1 state, abbreviated as (FE, AFM 1), the calculations of $\sigma^{z;zz}$ have been performed under different strain conditions, specifically, $\varepsilon = 0\%$, 2%, 3%, and 5% (Fig. 4(b)). For concise comparison, the frequency axes of the $\sigma^{z;zz}$ are rescaled by subtracting of the bandgap energy, $\omega - \omega_{\text{gap}} \to \omega$, where $\omega_{\text{gap}} = 0.79$ eV, 0.88 eV, 1.02 eV, and 1.22 eV. Without any strain ($\varepsilon = 0\%$), the shift current response vanishes due to the inversion symmetry, though finite dipole leads to a finite linear optical response [57]. Within the frequency range of $\omega = [0, 0.6]$ eV, the overall profiles of oscillatory shift current responses become larger with increasing frequency for different finite strains, and the increasing trend can be also inspected in linear optical response [57]. The enhancement in large frequencies is originated from the augmentation of transition strength $|r_{nm}(k)|^2$ among deep bands. As shown in the band structure (Fig. 5(a)) and $k$-resolved shift current response (Fig. 5c) under $\varepsilon = 3\%$, the near-gap shift current response mainly arise from the weak $C_{2v}$-symmetry-allowed transitions between the valence band maximum (VBM) featured with I $p_z$ orbitals at $\Gamma$ point and conduction band (I $p_z$ and W $(d_{x'z} + d_{y'z})$ orbitals). When the light frequency increases, the results of $k$-resolved shift current response indicate that the pronounced transitions at $k$ points of $\Gamma \to \pm Z$ gradually become the primary contribution (Fig. 5(c)), where the components of the flat valence bands turn to be W $d_{x'^2-y'^2}$ orbitals (Fig. 5(a)). Note that the $x'y'$-coordinates are marked in Fig. 2(a). As $\varepsilon$ increases, the flat valence bands are lifted close to the VBM (see Figs. S8-S11 in Ref. [57] ), and the profiles of the linear optical responses are brought forward [57]. However, within the frequency range of $\omega = [0, 0.6]$ eV, the shift current response under $\varepsilon = 3\%$ emerges as an optimal choice, reaching a maximum of about 16 $nm^2 \cdot \mu A/V^2$ at rescaled frequency of $\omega_{\text{max}} = 0.47$ eV. When yielding comparable strength of shift current responses, $\omega_{\text{max}}$'s of $\varepsilon = 2\%$ and 5% are larger than $\varepsilon = 3\%$. The effect of the FE distortion is then investigated from the shift current responses across changing ferroelectric states including results for PE, AFE, FE and FE$_{anti}$ states under $\varepsilon = 3\%$ (Fig. 4(b)). Here, the magnetic configuration of WOI$_3$ consistently follows the collinear AFM 1 order. In the FE$_{anti}$ state, flipping the direction of the FE polarization reverse the sign of $r_{nm;z}$, and the shift current response of FE$_{anti}$ state is opposite from FE state. By contrast, the shift current responses of PE and AFE states vanish due to inversion symmetry. Applying strains and coercive fields can be utilized to act as on-off and directional switches for NLO based devices.

The influence of the magnetic properties to shift current response is then explored by changing the magnetic configurations while maintaining the 1D WOI$_3$ in the FE state under $\varepsilon = 3\%$. When transitioning collinear AFM 1 to FM phase, the near-gap shift current response $\sigma^{z;zz}$ is shifted to higher light frequency (Fig. 4(d)). This altering of response arises from the change in $|r_{nm}(k)|^2$, as supported by additional information from $k$-resolved shift current response (Fig. 5(d)) and linear optical response (see Fig. S7b in Ref. [57]). The splitting of doubly degenerate bands noticeably reduces the densities of states (DOS) of the gap-edge energy regions (Figs. 5(a)-(b) and S12 in Ref. [57]), and optical selection rule dominates the suppression of transitions from the gap-edge valence band at the $k$ points of the valleys of conduction bands (Figs. 5(b)). Both effects then diminish the gap-edge transitions $|r_{nm}(k)|^2$ for FM order leading the altering of shift current response. Additionally, the strain-induced magnetic noncollinearity is found to have negligible effect onto the shift current response [57], while the changes for behaviors of shift current response are minor when altering the direction of the magnetic moments of WOI$_3$ from the easy axis to $a$ axis. It aligns with expectation that changing magnetic orders is more pronounced than fine-tuning magnetic configurations on the electronic structures.

## IV. Discussions

In this work, we report theoretical investigations on the 1D vdW WOI$_3$. A strain-induced paraelectric-ferroelectric phase transition is predicted with critical point of strain $\varepsilon_{CP} \sim 1.8\%$ for this material. Our calculations identify WOI$_3$ as a 1D ferroelectric-antiferromagnetic (multiferroic) semiconductor by strain engineering, where the magnetic moments are originated from the unpaired $d_{xy}$ electrons in $d^1$ configuration of the tungsten atoms. The strain-induced FE distortions (3% ≤ $\varepsilon$ ≤ 10%) can also lead 5°~12° noncollinear tilting of the magnetic moments, pristinely parallel to the easy axis ($\varepsilon = 0\%$), along the direction of the FE polarization. Subsequently, we explore the ferroelectric and magnetic properties concerning shift current BPVE in 1D WOI$_3$. Strains and electric fields serve as efficient controls for on-off and direction-switching of NLO responses. Additionally, changing the magnetic order has been identified as a way to manipulate NLO response. This opens the door for designing advanced optical devices with tailored properties.

We rationalize the multiferroicity in 1D vdW WOI$_3$ from the anisotropic $d^1$ rule, as proposed in works of 2D VOX$_2$(X=Cl, Br, I) [14,15]. The scarcity of systematic investigations into the anisotropic $d^1$ configuration in generating FE distortion and its implications in designing novel electromagnetic materials remains a significant gap in our understanding. This work sheds light on the discovery of new anisotropic $d^1$ materials, offering insights for the theoretical development of the anisotropic $d^1$ rule. Furthermore, 1D vdW WOI$_3$ is expected to be a multi-tune platform for devices utilizing shift current BPVE. The efficient control of shift current through the application of strains, external electric fields, and magnetic field is promising for potential applications in NLO devices. Due to the success in synthesis of the bulk vdW $WOI_3$ [35] and other similar structures [36–38], experimental verifications and applications in NLO devices could be achieved for the foreseeable future.

# Acknowledgements

H. W. and J. Q. acknowledge the support provided by National Natural Science Foundation of China (NSFC) under Grant No. 12304049 and 11974148. The work was carried out at National Supercomputer Center in Tianjin, and the calculations were performed on TianHe-1(A). Z. G. acknowledges Prof. Mingwen Zhao and Mr. Haoqiang Ai for useful discussions.

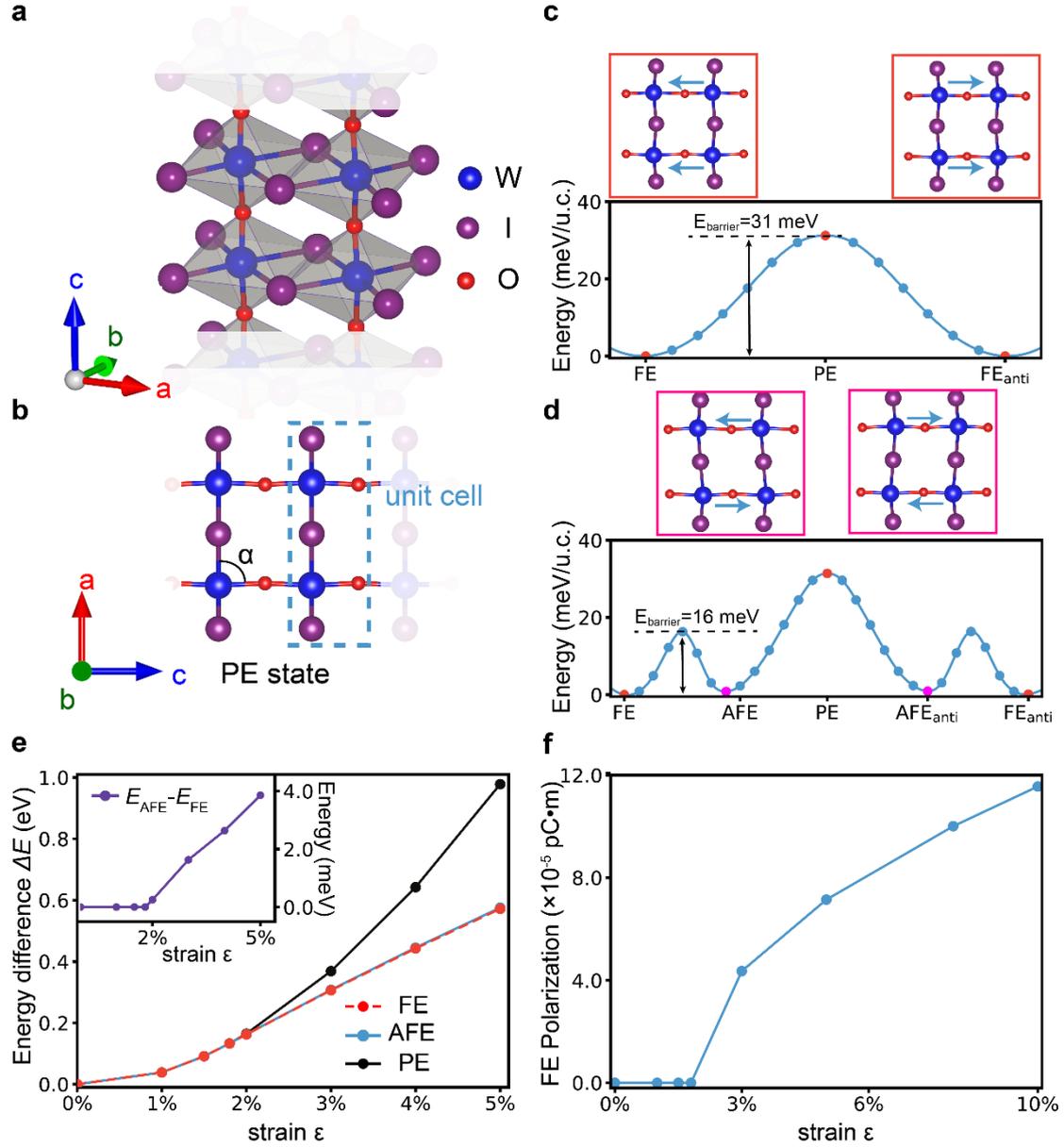

FIG. 1. Crystal structures and Ferroelectricity for the 1D WOI$_3$. The structures of PE state viewed in (a) the crystal orientation and in (b) the *b*-direction. The rung-like unit cell is highlight in the blue dashed box, and the specific O–W–I bond angle α in lower left corner of the cage of the $1 \times 1 \times 2$ cell characterizes the distortion degree. Under a uniaxial strain of $\varepsilon = 3\%$, the adiabatic energy pathways from the nudged elastic band (neb) are presented for (c) FE-PE-FE$_{anti}$ and (d) FE-AFE-PE-AFE$_{anti}$-FE$_{anti}$, with structures of the FE/FE$_{anti}$ and AFE/AFE$_{anti}$ states addressed. Here, the energy barriers of transition states are labelled. The FE$_{anti}$ represent the FE state with the opposite polarization direction. (e) The energy difference $\Delta E = E - E_{\text{PE},0\%}$ of different state under $\varepsilon = 0\%$-5%, and the energy difference between FE and AFE is presented in the inserted figure. (f) The FE polarization intensity along the periodic direction *c*-axis $P_c$ under different strains.

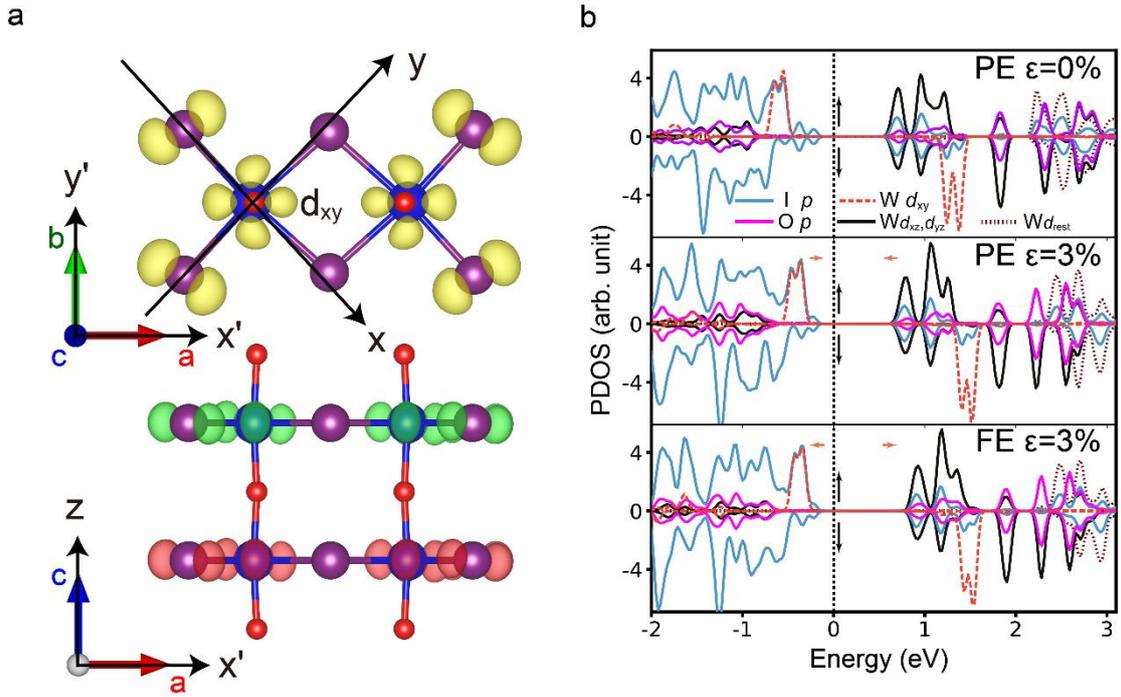

FIG. 2. Spatial charge densities and spin-polarized PDOS in supporting the ferroelectric anisotropic $d^1$ rule. (a) The charge densities represented by red (spin-up), green (spin-down) and yellow (both) isosurfaces with isosurface level of $3 \times 10^{-11}$ e/Bohr$^3$. The charge densities localized at W atoms correspond to the occupied $d_{xy}$ orbitals. Here, the $xy$-coordination axes for the orbitals are set to be aligned with the I-W bonds, distinguished from the marked $x'y'$-coordination axes. (b) The spin-up and spin-down PDOS for ($\varepsilon = 0\%$, PE), ($\varepsilon = 3\%$, PE), and ($\varepsilon = 3\%$, FE) are presented within the near-gap energy region. The red dashed and black solid lines represent the PDOS of $d_{xy}$ and $d_{xz} + d_{yz}$ orbitals for W atoms, while the brown dotted lines are PDOS for rest orbitals of W atoms. The p orbitals of I and O atoms are respectively denoted by blue and magenta solid lines.

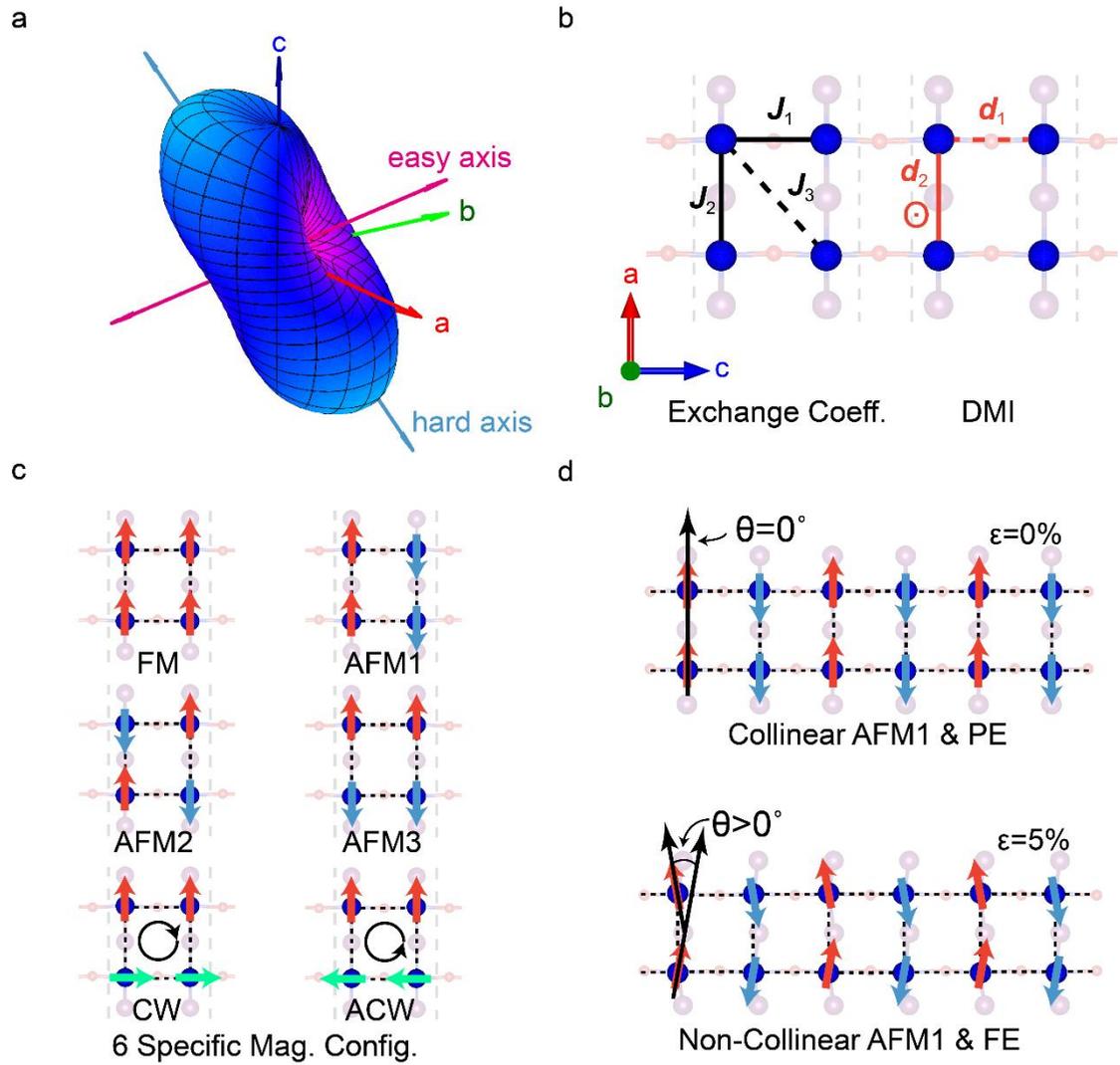

FIG. 3. MAE surface and schematics of Heisenberg models and magnetic configurations. (a) The MAE surface of the 1D WOI$_3$ with easy axis (red) and hard axis (blue). (b) The schematic diagrams for the exchange coefficients and DMI vectors considered in the anisotropic Heisenberg model. (c) six specific magnetic configurations used in determining parameters of the anisotropic Heisenberg model. These configurations are collinear Ferromagnetic (FM) order and three kinds of Antiferromagnetic (AFM 1-3) orders, and non-collinear Clockwise (CW) and Anticlockwise (ACW) configurations. (d) Schematic diagrams for the magnetic configurations under strain of ε=0% and ε=5%. The red and blue represent the magnetic moments at sites of W atom in a rung-like unit cell, and the averaged intersection angle θ is introduced to characterize the degrees of the noncollinearity.

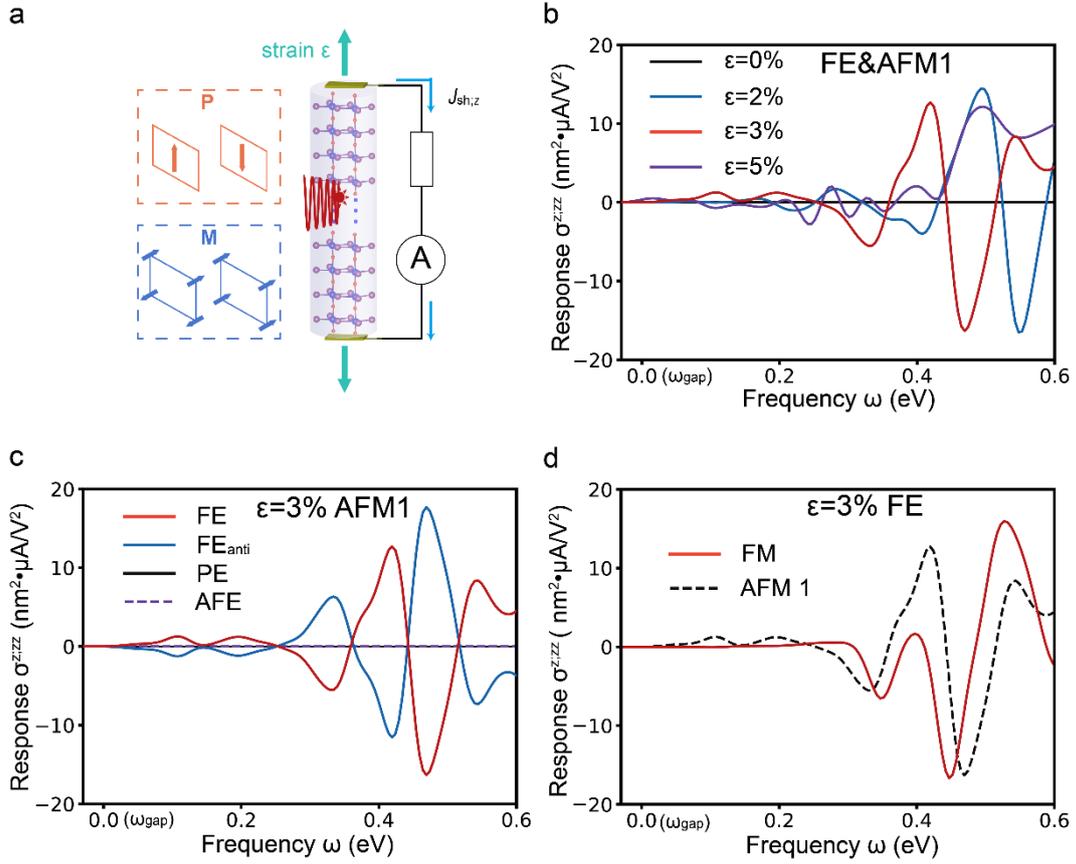

FIG. 4. Shift current response for 1D WOI$_3$. (a) The schematic diagram for the WOI$_3$-based NLO device. The shift current responses $\sigma^{z;zz}(\omega)$ for (b) (FE, AFM 1) state under $\varepsilon = 0\%$, 2%, 3%, and 5% represented by black, blue, red and purple lines; (c) FE, FE$_{anti}$, PE, and AFE structure with AFM 1 under $\varepsilon = 3\%$ represented by red, blue, black solid and purple dashed lines; (d) FE structure with AFM 1 and FM phases under $\varepsilon = 3\%$ denoted by black dashed and red solid lines. The collinear magnetic orientation is along with easy axis in (b)-(d).

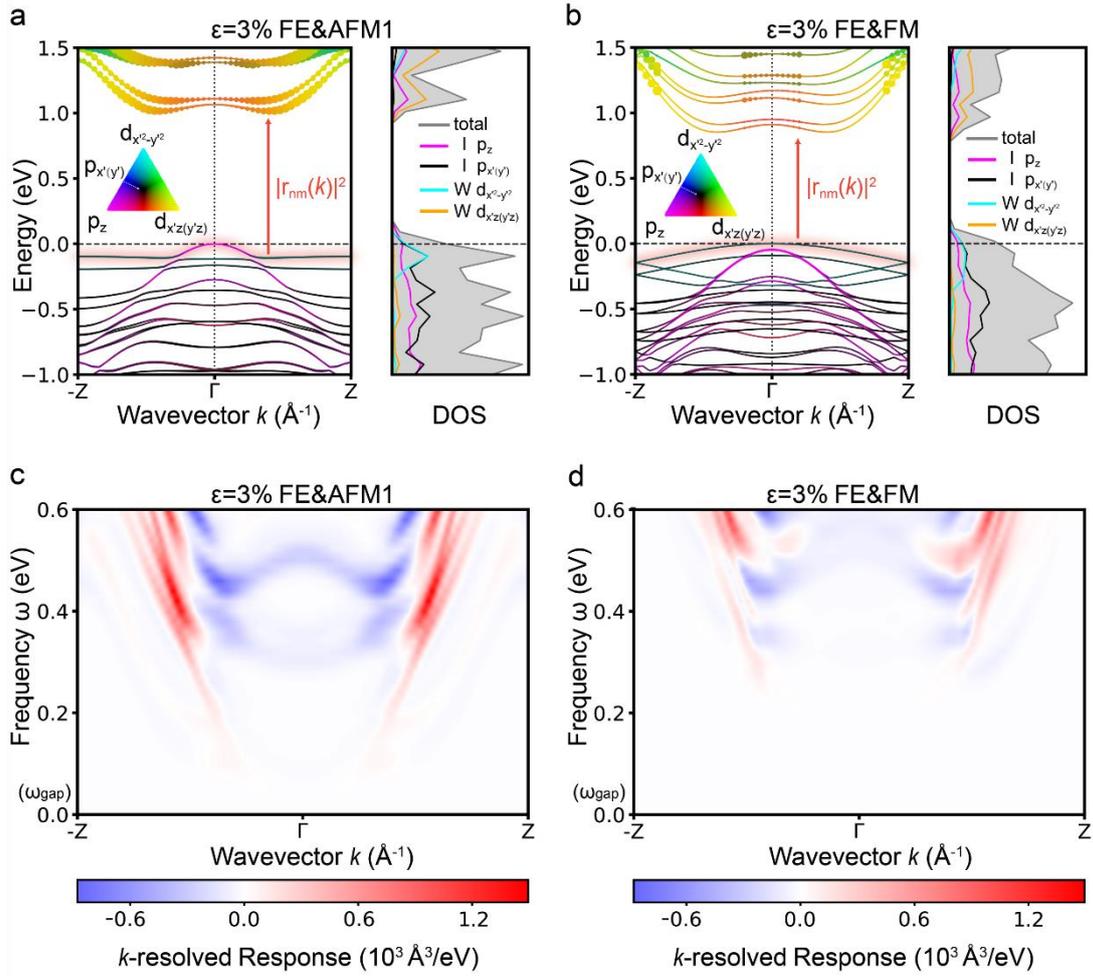

FIG. 5. Band structures and $k$-resolved shift current response for 1D $WOI_3$. The band structures and densities of states (DOS) of (a) (FE, AFM 1) and (b) (FE, FM) states under $\varepsilon = 3\%$. The color composition for each point of orbital projection in bands, using colors (yellow, cyan, black, and magenta), indicates the proportions of W $d_{x'z} + d_{y'z}$, W $d_{x'^2-y'^2}$, I $p_{x'} + p_{y'}$, and I $p_z$ orbitals. The sizes of solid circles represent the strength of transition dipole moments, $|r_{nm}(k)|^2$, from first gap-edge valence band to conduction bands. The total DOS and orbital PDOS are denoted by grey shaded line and colored lines, respectively. The $k$-resolved shift current responses of (c) (FE, AFM 1) and (d) (FE, FM) states under $\varepsilon = 3\%$ are plotted with $k$ points and light frequencies.

TABLE I. The energy difference, $\Delta E = E - E_{\text{PE},0\%}$, (left panel) and the angle, $\alpha$, characterizing the magnitude of structure distortion (right panel) for PE, FE, and AFE states under strain $\varepsilon = 0\%$ to $5\%$.

| STRAIN | $\Delta E$ (eV) | | | $\alpha$ (°) | | |
|---|---|---|---|---|---|---|
| $\varepsilon$ | PE | FE | AFE | PE | FE | AFE |
| **0%** | 0 | – | – | 88.0972 | – | – |
| **1%** | 0.038583 | 0.038583 | 0.038583 | 88.1990 | 88.2007 | 88.2046 |
| **1.5%** | 0.091284 | 0.091285 | 0.091288 | 88.2475 | 88.2120 | 88.2250 |
| **1.8%** | 0.132926 | 0.132926 | 0.132927 | 88.2725 | 88.2360 | 88.2506 |
| **2%** | 0.164657 | 0.162495 | 0.162743 | 88.2942 | 86.8346 | 89.6263 |
| **3%** | 0.368723 | 0.306299 | 0.307929 | 88.4168 | 84.7014 | 91.5060 |
| **4%** | 0.642511 | 0.442843 | 0.445473 | 88.4168 | 83.7014 | 92.6567 |
| **5%** | 0.978206 | 0.571920 | 0.575775 | 88.4676 | 82.7923 | 93.5199 |

TABLE II. The results of NN exchange coefficients and DMI, $J_1, J_2$, and $d_{2;b}$, for the anisotropic Heisenberg model and the averaged intersection angle $\theta$ under strains of $\varepsilon = 0\%$ to $10\%$

| STRAIN $\varepsilon$ | | $J_1$ | $J_2$ | $d_{2;a=b}$ | $\theta$ (°) |
|---|---|---|---|---|---|
| | | | (meV/$\mu_B^2$) | | |
| 0% | a: | 1.07(2) | −4.28(8) | 0 | 0 |
| | b: | 0.79(4) | −4.50(9) | | |
| | c: | 1.18(6) | −4.44(5) | | |
| 3% | a: | 0.85(9) | −3.98(1) | 0.49 (5) | 5.69(2) |
| | b: | 0.63(9) | −3.81(5) | | |
| | c: | 0.95(6) | −4.13(2) | | |
| 5% | a: | 0.72(5) | −3.77(3) | 0.62(3) | 7.36(5) |
| | b: | 0.60(5) | −3.15(9) | | |
| | c: | 0.77(4) | −3.92(6) | | |
| 8% | a: | 0.49(8) | −3.57(7) | 0.91(9) | 10.91(8) |
| | b: | 0.45(8) | −2.66(8) | | |
| | c: | 0.51(1) | −3.75(7) | | |
| 10% | a: | 0.40(4) | −3.58(3) | 0.98(0) | 11.68(6) |
| | b: | 0.37(1) | −2.59(6) | | |
| | c: | 0.41(5) | −3.76(9) | | |